\documentclass[letterpaper, 10 pt, conference]{ieeeconf}  
\IEEEoverridecommandlockouts                              

\usepackage{soul}
\usepackage{multirow}
\usepackage[normalem]{ulem} %
\usepackage{subcaption} 
\usepackage{graphicx}
\usepackage{dblfloatfix}    %

\usepackage{color}
\newcommand{\edit}[1]{\textbf{\textcolor{red}{#1}}}
\newcommand{\chou}[1]{\textcolor{blue}{#1}}

\title{\LARGE \bf
Improved Optical Flow for Gesture-based Human-robot Interaction
}

\author{
    Jen-Yen Chang$^{1}$, Antonio Tejero-de-Pablos$^{1}$ and Tatsuya Harada$^{12}$%
\thanks{$^{1}$ is with the Graduate School of Information Science and Technology, The University of Tokyo, Japan. Email: \{chou, antonio-t, harada\}@mi.t.u-tokyo.ac.jp}
\thanks{$^{2}$ is with the RIKEN Center for Advanced Intelligence Project (RIKEN AIP), Tokyo, Japan. Email: tatsuya.harada@riken.jp}
}

\begin{document}

\maketitle
\thispagestyle{empty}
\pagestyle{empty}

\begin{abstract}

Gesture interaction is a natural way of communicating with a robot as an alternative to speech. Gesture recognition methods leverage optical flow in order to understand human motion. However, while accurate optical flow estimation (i.e., traditional) methods are costly in terms of runtime, fast estimation (i.e., deep learning) methods' accuracy can be improved.
In this paper, we present a pipeline for gesture-based human-robot interaction that uses a novel optical flow estimation method in order to achieve an improved speed-accuracy trade-off.
Our optical flow estimation method introduces four improvements to previous deep learning-based methods: strong feature extractors, attention to contours, midway features, and a combination of these three.
This results in a better understanding of motion, and a finer representation of silhouettes.
In order to evaluate our pipeline, we generated our own dataset, MIBURI, which contains gestures to command a house service robot.
In our experiments, we show how our method improves not only optical flow estimation, but also gesture recognition, offering a speed-accuracy trade-off more realistic for practical robot applications.

\end{abstract}

\section{Introduction}
\label{sec:I}
In our society, people have a constant need to communicate with each other (e.g., via voice or gestures). Although voice is normally preferred for its directness, sometimes verbal communication is not possible due to environmental factors such as noise, or the inability of one of the interlocutors to communicate, due to, e.g., a speech/hearing disability or when the interlocutor speaks in another language or has a very strong accent. In such cases, humans use gestures as an alternative modality for interaction \cite{c21}.
On the other hand, robots are increasingly being adopted in our daily life. There has been a rising demand for domestic service robots for entertainment purposes \cite{so}, taking care of elders or hospital patients \cite{longterm}, and carrying out domestic tasks. An example of these robots are the Human Support Robot (HSR) developed by Toyota and Pepper developed by Softbank Robotics. The speech recognition modules installed in house service robots (e.g., DeepSpeech \cite{DeepSpeech} or Google Speech API) have achieved near human level performance. However, similarly to humans, human-robot verbal communication is also affected by factors such as noisy environments, a malfunctioning microphone, or situations in which the robot and the human do not ``speak'' the same language. In order to achieve a more reliable human-robot interaction (HRI), an alternative way to communicate with the robot is necessary. As with humans, gestures provide an alternative modality for communication, and thus, gesture recognition plays an important role in HRI.

\begin{figure}[t]
    \caption{Overview of our pipeline. The input video from the robot's camera is used to estimate the optical flow with our novel method. Then, through optical flow-based gesture recognition, the robot's operating system (e.g., ROS) selects which module to run. }
    \centering
    \includegraphics[width=0.9\linewidth]{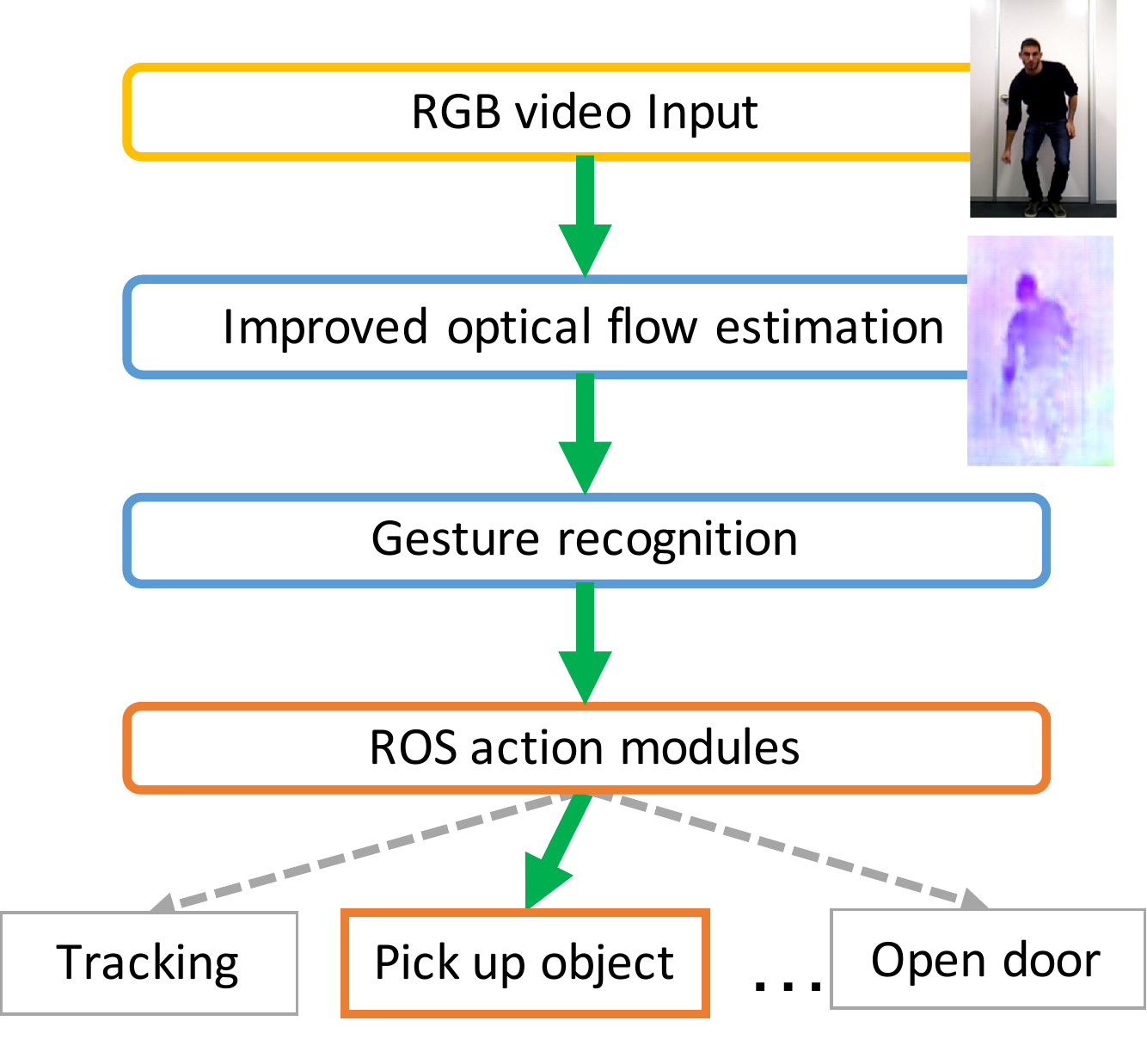}
    \label{fig:pipeline}
\end{figure}

In computer vision, gestures can be recognized using an action recognition method \cite{Simonyan} \cite{TSN}. These methods include optical flow estimation in their pipeline to facilitate understanding the dynamic nature and the spatio-temporal features of gestures. Current optical flow estimation methods have limitations: (1) they result in a blurry representation of the human silhouette that misses a lot of detail, and (2) they are time-costly. Algorithms with an excessive runtime are not suitable for real time gesture-based HRI. If recognizing a gesture takes too long (e.g., 30s), communication is heavily hindered. On the other hand, if the recognition time is fast but the accuracy is poor, recognition mistakes will make the communication impossible.

In this work, we present a gesture recognition pipeline (Fig. \ref{fig:pipeline}) designed for human-robot interaction that achieves a good trade-off balance between speed and accuracy. In order to obtain a finer spatio-temporal representation of human motion, we proposed a novel optical flow that pays attention to the human body silhouette, and thus, achieves a better recognition accuracy. Our contributions are:
\begin{itemize}
    \item Three novel optical flow estimation methods that incorporate attention to allow for better gesture recognition.
    \item An optical-flow based gesture recognition pipeline with improved speed-accuracy trade-off, designed for human-robot interaction.
    \item A self-generated dataset for human-robot interaction, MIBURI\footnote{MIBURI project URL: https://www.mi.t.u-tokyo.ac.jp/projects/miburi/}, designed to command the robot to complete household tasks.
\end{itemize}

\section{Related work}

\subsection{Human-Robot Interaction Via Gestures}
Nowadays, most human-robot interaction methods are speech recognition-based \cite{c01} \cite{hri}. Unfortunately, the use of speech recognition is somewhat limited, due to environmental, hardware, or linguistic problems (Sec. \ref{sec:I}). Gesture communication serves as a natural alternative to speech \cite{c21}.
Previous works \cite{generative, industry} study the effect of robots performing gestures to communicate with humans. However, our interest is in humans commanding the robot through gestures.
Most existing works on human-robot interaction (HRI) via gestures focus on hand gesture recognition \cite{gesture}. However, hand gestures limit the motion to only the hand part of the human body, which may not be intuitive enough for general human users to perform. 
Unlike these works, we also refer to motions involving the whole body as gestures.

\subsection{Gesture Recognition and Optical-Flow based Gesture Recognition}
\label{sec:II-B}

In robot vision, gestures can be recognized using action recognition methods that combine appearance and motion.
Whereas color-only methods \cite{UCF101} can recognize certain activities such as sports, their representation of motion is not fine enough to differentiate between similar gestures. In contrast, optical flow is a better representation of motion.
Deep learning-based action recognition using optical flow achieved outstanding results with the two-stream network approach \cite{Simonyan} developed by Simonyan et al.. They trained two separate convolutional neural networks to explicitly capture spatial features from color images and temporal features from optical flow. Finding Action Tubes \cite{Tubes} extended the work of Simonyan et al. by proposing candidate regions descriptors instead of full images in order to localize the action to improve the recognition accuracy. However, the trade-off between the runtime of recognizing the gesture and accuracy of such recognition can be improved. Recent approaches such as Temporal Segment Networks (TSN) \cite{TSN}, and \cite{Two-Stream} further extended the two-stream approach. More concretely, TSN proposed to concatenate video frames into segments for a better speed-accuracy trade-off.

\subsection{Optical Flow Estimation}
\label{sec:II-C}
In optical flow-based gesture recognition, such as the previously described \cite{Two-Stream} and \cite{TSN}, in order to estimate optical flow from the input images, researchers deployed traditional optical flow estimation methods, namely the TVL1 \cite{TVL1}, and the work of Brox et al. \cite{Brox}. These optical flow estimation methods reached state-of-the-art accuracy, but they are slow. To tackle this, researchers leveraged deep learning techniques. The first attempt of a deep optical flow estimation network, FlowNet \cite{FlowNet}, proposed two networks, namely FlowNetS and FlowNetC, and greatly improved the runtime of previous work. Zhu et al. \cite{DenseF} developed an optical flow estimation method called DenseNet by extending a deep network for image classification, and demonstrated that it outperforms other unsupervised methods. FlowNet 2.0 \cite{FlowNet2} is a follow-up paper of FlowNet \cite{FlowNet}. They stacked two optical flow estimation networks (FlowNetS and FlowNetC) and achieved state-of-the-art results comparable to traditional optical flow methods while running on 8 to 140 fps depending on the desired accuracy (the higher the accuracy, the slower the runtime). However, one of the main problems of FlowNet 2.0 \cite{FlowNet2} is its model size (parameter counts). Since the architecture is stacked, the parameter count is naturally high. Recently, some methods focused on obtaining the same accuracy while reducing its model size \cite{MSCSL}.

\section{Optical Flow-based Gesture Recognition}

Figure \ref{fig:pipeline} depicts an overview of the basic pipeline of our gesture recognition method for human-robot interaction (HRI). The RGB video of the user performing a gesture is fed to our optical-flow estimation method, and then, the resulting optical flow is used to recognize the gesture.

We implemented our modules by taking into account the trade-off between the recognition accuracy and speed.
While most gesture/action recognition methods employ optical flow (Sec. \ref{sec:II-C}), they mainly use traditional optical flow estimation. We argue that using deep learning for estimating optical flow results in a better trade-off between gesture recognition runtime and accuracy, and thus, is adequate for HRI. We also argue that adding an attention mechanism to deep learning optical flow estimation results in a higher accuracy, not only for optical flow estimation (reduces blurriness in contours), but also for gesture recognition (refines the user's silhouette).

Following the same criteria, we feed our optical flow to the TSN action recognition network \cite{TSN} (Sec. \ref{sec:II-B}), as it features state-of-the-art accuracy with short recognition time. That is, approximately one second on a GTX 1080 GPU, for action recognition on 120 images (4 seconds of video at 30fps).

The next section explains in detail our contribution in optical flow estimation.

\begin{figure*}[!t]
  \caption{Proposed optical flow estimation networks. Top left: original FlowNetS \cite{FlowNet}. Top right: AttFlowNet\{Res, Inc, NeXt32, NeXt64\}. Bottom left: MidFlowNet\{Res, Inc, NeXt32, NeXt64\} \cite{FlowNet}. Bottom right: AttMidFlowNet\{Res, Inc, NeXt32, NeXt64\}. For all figures below, orange blocks are convolutional layers (feature extractors), blue blocks are transposed convolutional layers, green blocks are optical flow estimators, yellow blocks are attention calculators. Numbers indicate layer size (same number, same size).}
  \begin{minipage}[b]{\textwidth}
  \centering
    \includegraphics[height=4.5cm, width=0.5\linewidth, keepaspectratio]{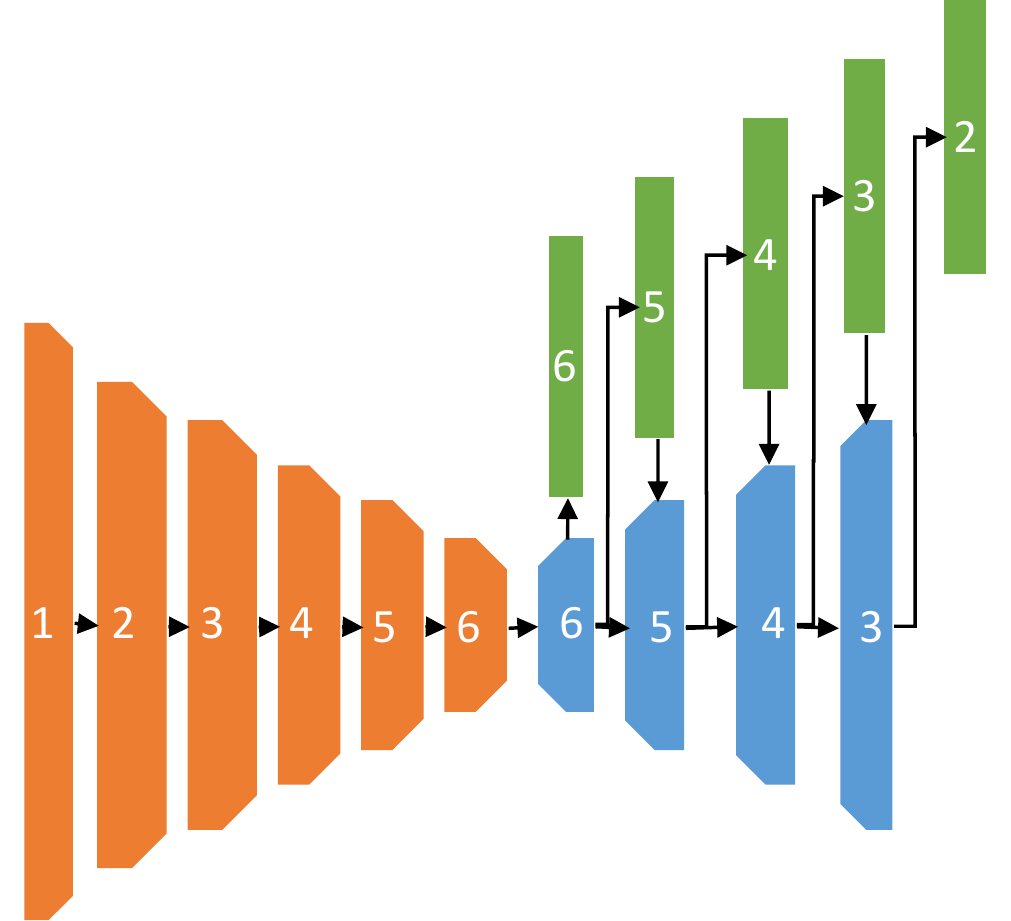}
    \includegraphics[height=4.5cm, width=0.5\linewidth, keepaspectratio]{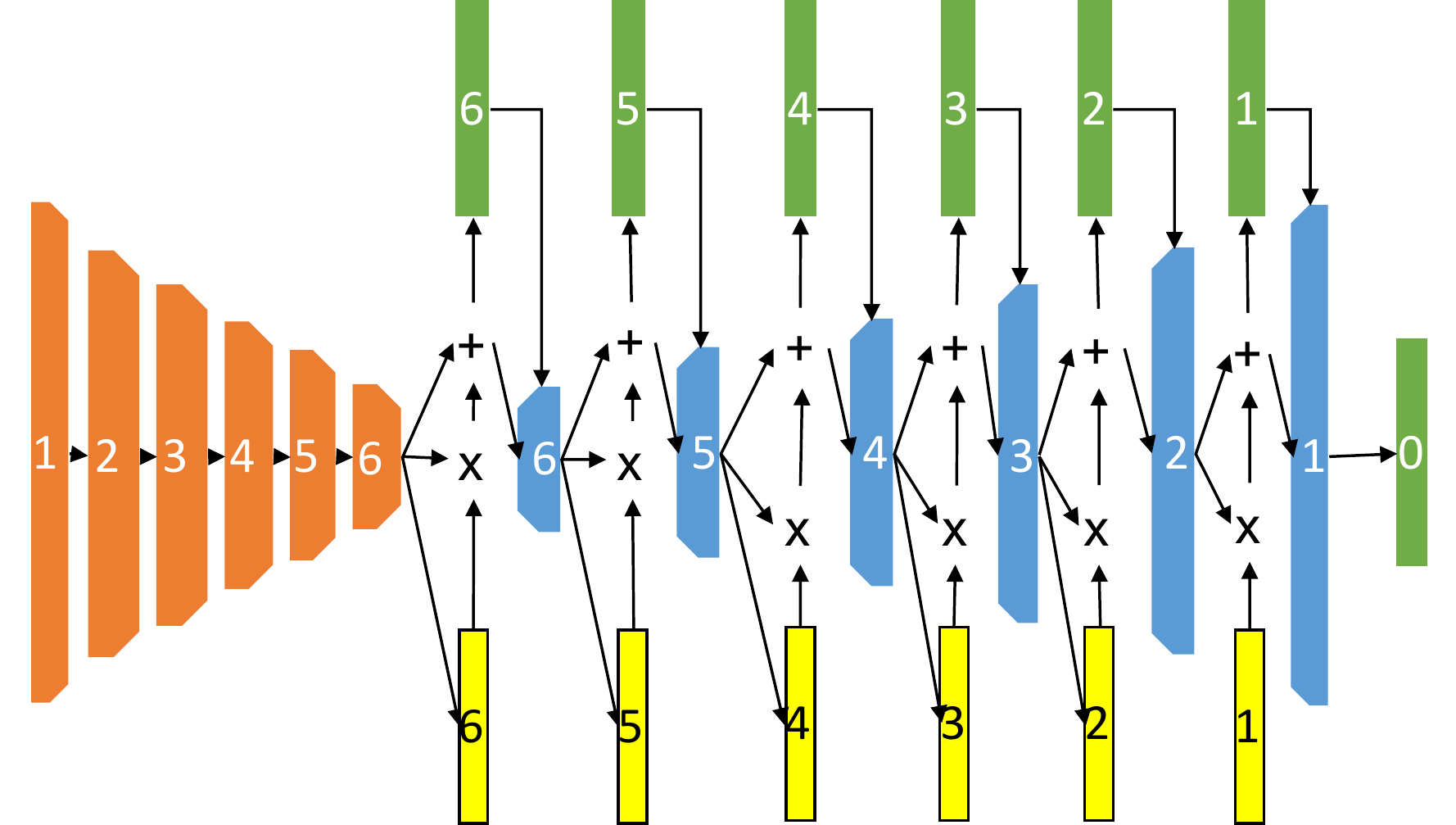}
  \end{minipage}
  \hfill \\
  \begin{minipage}[b]{\textwidth}
  \centering
    \includegraphics[height=5cm, width=0.5\linewidth, keepaspectratio]{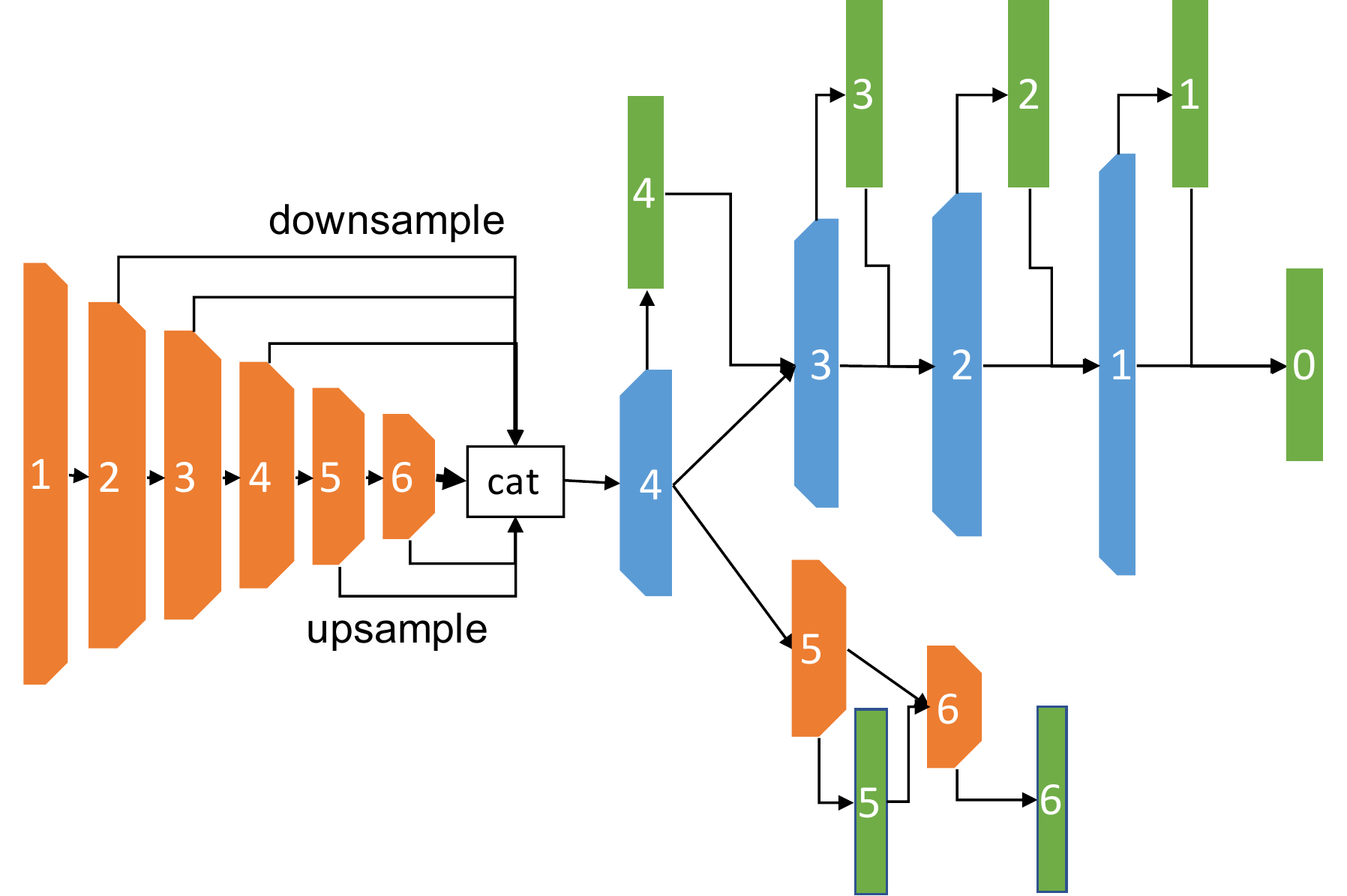}
    \includegraphics[height=5cm, width=0.5\linewidth, keepaspectratio]{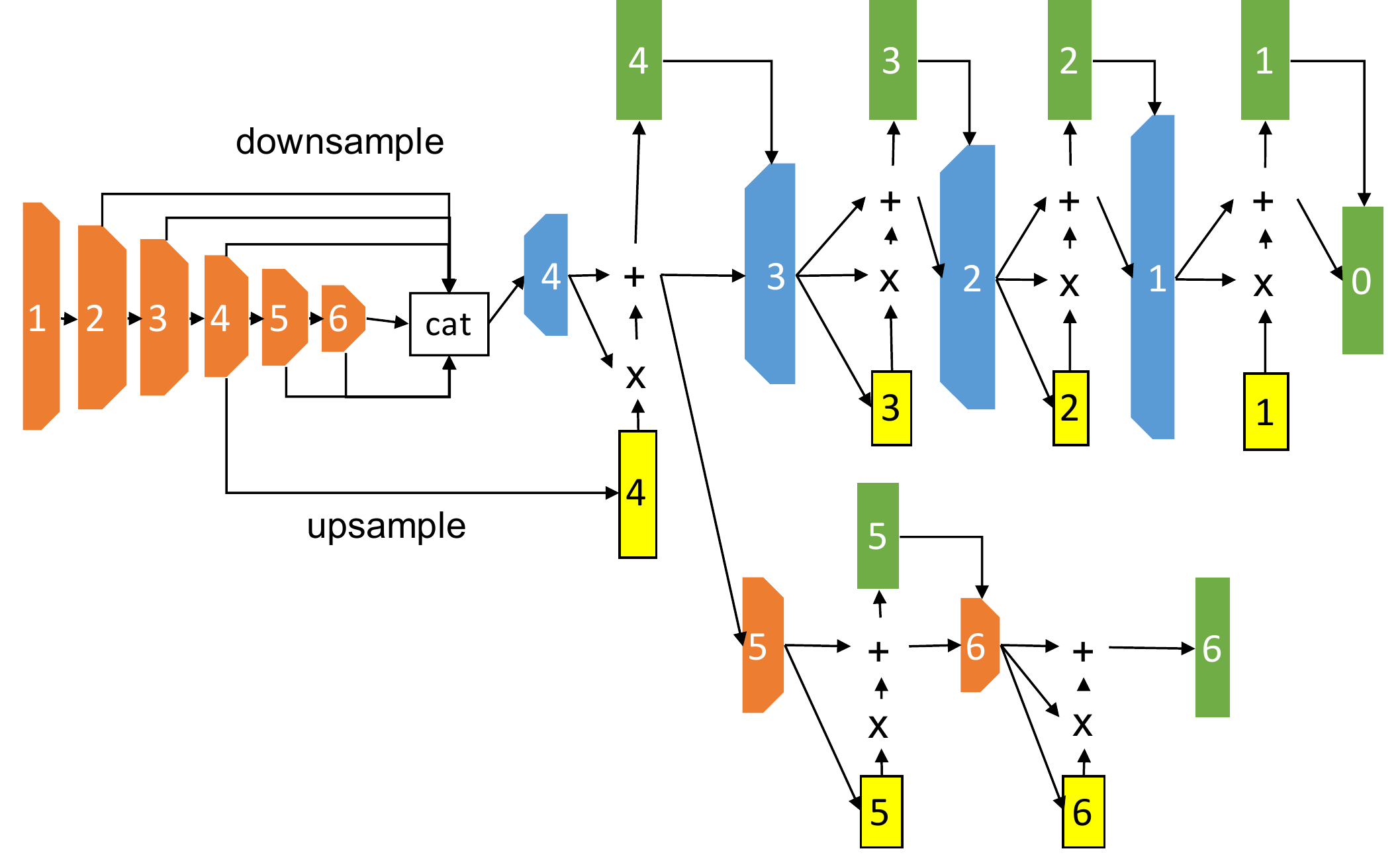}
  \end{minipage}
  \label{fig:1}
\end{figure*}

\section{Optical flow estimation}
We define optical flow estimation as the task of predicting motion pixel-wise in two consecutive frames. It can be also seen as a way of understanding the spatio-temporal features of movements, actions and gestures. We aim to improve estimation accuracy in three ways: by using stronger image feature extractors, by adding attention to the estimation, and by emphasizing mid-level feature representations. Based on this, we propose three novel optical flow estimation networks.

\subsection{Stronger feature extractors}
Optical flow estimation aims to predict how pixels move, and thus, we have to understand images at pixel-level. That is, first we extract features from each pixel of an image, then we match the features between two consecutive images, and finally we estimate pixel-level movement. Therefore, we can expect that a better understanding of the images results in a more accurate matching. As shown in \cite{DenseF}, in optical flow estimation, feature extractors trained on image classification tasks are recommendable for unsupervised learning. 

Our novel optical flow estimation uses stronger feature extraction blocks (i.e. blocks capable of extracting more discriminative features, tested in image classification tasks) onto an existing baseline estimation network. We chose FlowNetS \cite{FlowNet} as our base network due to its well-balanced speed-accuracy trade-off. For this purpose, we resorted to four feature extractors widely used for classifying the ImageNet dataset \cite{imagenet}: ResNet \cite{resnet}, Inception \cite{inception}, ResNext (4x32d block) and ResNext (4x64d block) \cite{resnext}. We replaced the simple convolutional layers present in FlowNetS \cite{FlowNet} with an adaptation of ResNet, Inception and ResNext. We named these optical flow estimation networks as FlowNetRes, FlowNetInc, FlowNeXt32 and FlowNeXt64 respectively (Fig. \ref{fig:1}).

\subsection{Attention module}

Once the feature extraction is improved, we target the optical-flow estimation.
Ideally, pixels belonging to the same entity (e.g., a hand) should have similar optical flow values. This makes the contours between different entities sharper. To achieve this, our novel network adopts an attention mechanism that was originally used for putting emphasis on keywords in the field of natural language processing \cite{NLP}. This attention mechanism enlarges the difference between dissimilar optical flow values, so estimations for pixels within the same moving entity will be similar. The overall formula for applying attention to a layer is
$$
y = sigmoid(x + x * attention(x)), \eqno{(1)}
$$
where $x$ is the input features to the layer, $y$ is the output of the layer, and $attention(x)$ is a set of convolutional layers (details in Table \ref{table:attention}). When $attention(x)$ is 0, the output features will be the same as the original features. Finally, a sigmoid activation function is applied. Thus, large estimations are enlarged whereas smaller estimations remain similar. Attention allows preserving the silhouettes of the moving parts of the image. Figure \ref{fig:1} (top-right) shows a schematic diagram, derived from the original FlowNetS \cite{FlowNet}. We named these optical flow estimation networks as AttFlowNetRes, AttFlowNetInc, AttFlowNext32, and AttFlowNext64 respectively. It is also noted that starting from here, we added additional upscaling back to the original resolution, as the final estimation resolution of FlowNetS \cite{FlowNet} is only to the 1/4 of the original resolution.

\begin{table}[t]
  \centering\fontsize{8}{10}\selectfont
  \smallskip\caption{Architecture of our attention module}
  \begin{tabular}{|c|cc|c|c|}
  \hline
  Name & Kernel & Strides & Input & Ch I | O        \\	\hline\hline
  attconv1   & 3x3 & 2 & x           &    x | x/4 \\
  attconv2   & 3x3 & 2 & attconv1    &  x/4 | x/8  \\
  attdeconv1 & 4x4 & 2 & attconv2    &  x/8 | x/4  \\
  attdeconv2 & 4x4 & 2 & attdeconv2  &  x/4 | x    \\ \hline
  \end{tabular}
  \label{table:attention}
\end{table}

\subsection{Midway module}
One of the major challenges with optical flow estimation is predicting a correct general direction of the motion. This is normally solved by using a pyramid structure, like the one presented in SpyNet \cite{spynet} or \cite{LK}. These pyramid methods learn features starting from the largest feature patches down to the smallest ones. However, we lose estimation accuracy when we take the same strategy on estimations when we upscale back to the original resolution. We argue that, in order to estimate the overall optical flow more accurately, mid-level estimations are crucial. Mid-level means that the scale of the estimation is scaled-down from the original input frame size, but not too small so that each pixel can represent its nearby estimation but will not over represent. Therefore, the estimation will have more uniform patches overall, and no sharp changes inside the same object, but will have sharp contrasts especially on the contours. Therefore, we designed a novel module that puts emphasis on the mid-level estimations. A schematic diagram is shown in Fig. \ref{fig:1} (bottom-left). We named these optical flow estimation networks MidFlowNetRes, MidFlowNetInc, MidFlowNext32, and MidFlowNext64 respectively. Table \ref{table:midway} shows the architecture and layer sizes of our MidFlowNet.

\begin{table}[ht]
  \centering\fontsize{8}{10}\selectfont
  \smallskip\caption{General architecture of MidFlowNet \{Res,Inc,NeXt\}. Bi\_up means bilinear upsampling. Avg\_down means average-pooling. the output of biup and avgdown has the same size as block4\_1. Pr means optical flow estimation. Loss is EPE loss.  Deconv is deconvolution followed by LeakyRelu activation. Up is only deconvolution. Concat concatenates at the channel axis. }
  \begin{tabular}{|c|cc|c|c|}
  \hline
  Layer Name & Kernel & Strides & Input & Ch I/O    	\\	\hline\hline
  block1  	& 7x7 & 2 & Im1+Im2    &    6/64   	\\
  block2  	& 5x5 & 2 & block1     &   64/128 	\\
  block3  	& 5x5 & 2 & block2     &  128/256  	\\
  block3\_1 & 1x1 & 1 & block3     &  256/256   \\
  block4    & 1x1 & 2 & block3\_1  &  256/512  	\\
  block4\_1 & 1x1 & 1 & block4     &  512/512   \\
  block5    & 1x1 & 2 & block4\_1  &  512/512   \\
  block5\_1 & 1x1 & 1 & block5     &  512/512  	\\
  block6    & 1x1 & 2 & block5\_1  &  512/1024  \\
  block6\_1 & 1x1 & 1 & block6     & 1024/1024  \\	\hline
  biup\_1    &&       & block6\_1  & 1024/1024  \\
  biup\_2    &&       & block5\_1  &  512/512   \\
  avgdown\_1 &&       & block3\_1  &  256/256   \\
  avgdown\_2 &&       & block2     &  128/128   \\  \hline
  
  \multirow{2}{*}{concat}&\multicolumn{3}{|c|}{biup\_1 + biup\_2 + block4\_1} 
  & \multirow{2}{*}{-/2432}	\\ 	 
  & \multicolumn{3}{|c|}{+ avgdown\_1 + avgdown\_2} & \\ 	\hline	
  pr4+loss4 & 3x3 & 1  & concat		                        & 2432/2   	\\ \hline
  down4     & 4x4 & 2  & pr4+loss4	                        & 	 2/2    \\
  deconv4   & 4x4 & 2  & concat 	                        & 2432/512  \\ 	\hline
  concat5&\multicolumn{3}{|c|}{block5\_1 + deconv4 + down4} &    -/1026 \\ 	\hline
  pr5+loss5 & 3x3 & 1  & concat5 				            & 1026/2 	\\  \hline
  
  down5     & 4x4 & 2  & pr5+loss5	                        &	 2/2    \\
  deconv5   & 4x4 & 2  & concat5                            & 1026/1024 \\ 	\hline
  concat6&\multicolumn{3}{|c|}{block6\_1 + deconv5 + down5} &    -/2050 \\ 	\hline
  pr6+loss6 & 3x3 & 1  & block6\_1                          & 2050/2    \\  \hline
  
  up4       & 4x4 & 2  & pr4+loss4				            &    2/2   	\\
  deconv3   & 4x4 & 2  & concat 				            & 2432/256 	\\ 	\hline
  concat3&\multicolumn{3}{|c|}{block3\_1 + deconv3 + up4}   &    -/514	\\ 	\hline
  pr3+loss3 & 3x3 & 1  & concat3				            &  514/2    \\  \hline
  
  up3       & 4x4 & 2  & pr3+loss3				            &	2/2     \\
  deconv2   & 4x4 & 2  & concat3 				            & 512/128	\\ 	\hline
  concat2&\multicolumn{3}{|c|}{block3\_1 + deconv2 + up3}	&   -/258 	\\ 	\hline
  pr2+loss2 & 3x3 & 1  & concat2 				            & 258/2	  	\\ 	\hline
  
  up2       & 4x4 & 2  & pr2+loss2				            &	2/2     \\
  deconv1   & 4x4 & 2  & concat2 				            & 258/64	\\ 	\hline
  concat1&\multicolumn{3}{|c|}{deconv1 + up2}	            &   -/66 	\\ 	\hline
  pr1+loss1 & 3x3 & 1  & concat1				            &  66/2	  	\\ 	\hline
  
  up1       & 4x4 & 2  & pr1+loss1				            &	2/2     \\
  deconv0   & 4x4 & 2  & concat1 				            &  66/32	\\ 	\hline
  concat0&\multicolumn{3}{|c|}{deconv0 + up1}	            &   -/34 	\\ 	\hline
  pr0+loss0 & 3x3 & 1  & concat0 				            &  34/2	  	\\ 	\hline
  
  \end{tabular}
  \label{table:midway}
\end{table}

\subsection{Attention-midway module}
Our proposed novel optical-flow estimation method is based on the combination of attention and midway. This way, we manage to simultaneously put emphasis on both the mid-level estimations and the contours. Figure \ref{fig:1} shows a schematic diagram. We named these optical flow estimation networks as AttMidFlowNetRes, AttMidFlowNetInc, AttMidFlowNext32, and AttMidFlowNext64 respectively.

\section{Experiments}

\label{sec:V}
To the best of our knowledge, previous gesture recognition works do not discuss whether having a more accurate optical flow estimator results on more accurate gesture recognition. In order to shed light on the matter, we have conducted two experiments. First, we compared our optical flow estimation methods against the state of the art. Then, we evaluate the improvement in gesture recognition accuracy when using our estimated optical flow. In each experiment, our proposed method for optical flow estimation and gesture recognition is trained end-to-end. For this, in order to evaluate the recognition of gestures to command a home service robot, we generated our own dataset, MIBURI. The runtime provided in this section corresponds to the execution times of our algorithms in an Nvidia TITAN X GPU, with data stored on SSD (to minimize data reading time).

\subsection{MIBURI: Video gesture Dataset for commanding Home Service Robots}

In order to evaluate the performance of gesture recognition for commanding a home service robot, a video dataset of gestures is required. Since, to the best of our knowledge, no existing public dataset fits our needs, we generated our own.

\begin{figure}
  \centering
  \caption{Examples of our MIBURI dataset (top): pick-bag, drop-bag, on-the-left. Examples of the ChaLearn IGR dataset (bottom): surgeon needle holder, italian d'accordo, volleyball referee substitution}
  \begin{minipage}[b]{\textwidth}
    \includegraphics[width=0.15\linewidth]{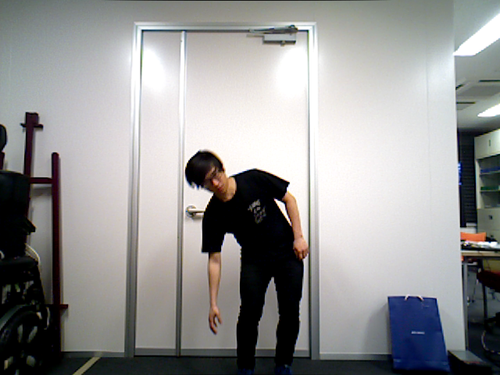}
    \includegraphics[width=0.15\linewidth]{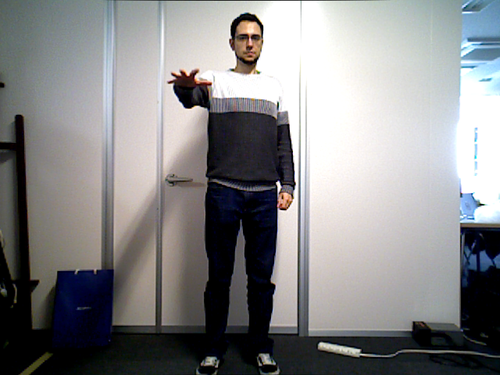}
    \includegraphics[width=0.15\linewidth]{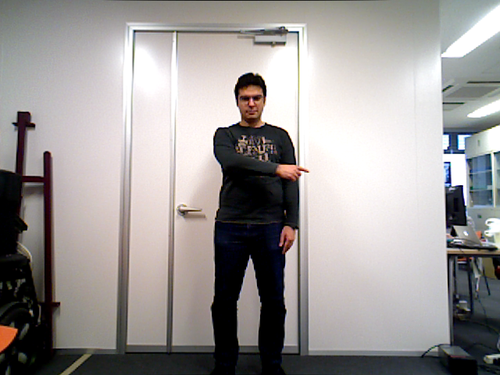}\newline\newline
    \includegraphics[height=0.12\linewidth, width=0.15\linewidth]{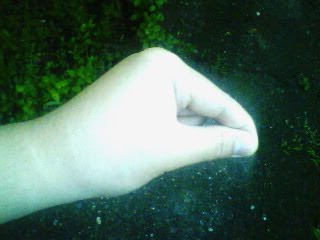}
    \includegraphics[height=0.12\linewidth,width=0.15\linewidth]{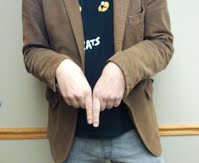}
    \includegraphics[height=0.12\linewidth,width=0.15\linewidth]{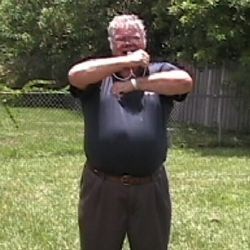}
  \end{minipage}

  \label{fig:miburi}
\end{figure}

Our dataset, MIBURI, was collected using HSR robot's head camera, an ASUS Xtion (640$\times$480 resolution). It contains 10 gestures: \textit{pick-bag}, \textit{drop-bag}, \textit{follow-me}, \textit{stop-follow-me}, \textit{bring-drink}, \textit{bring-food}, \textit{on-the-left}, \textit{on-the-right}, \textit{yes/confirm}, \textit{no/decline} (see examples in Fig. \ref{fig:miburi}).

When selecting the motion for each gesture, we applied the following criteria: (1) they are easily understandable for general people, even the first time they see them; (2) they are natural for humans to perform. These two points are essential from the perspective of HRI, since gestures should be used effortlessly to command the robot \cite{hand}. We recorded gestures for 15 users (among them 3 females and 12 males ranged from age 22 to age 30) and four repetitions per gesture, which sums up to a total of 600 videos consisting of 69835 frames. The average duration of the video is 3.88s.

\begin{table}[t]
  \setlength{\tabcolsep}{5pt}
  \centering\fontsize{8}{10}\selectfont
  \caption{Results of optical flow estimation in EPE (End Point End, the lower the better). The runtime reported is on FlyingChairs.}
  \begin{tabular}{|c||c|c|c|c|c|}
  \hline
  \multirow{3}{*}{Model} & Flying- & Sintel & Sintel  & & Flow \\
  & Chairs & \textit{clean} & \textit{final} & Params. & runtime \\
  & (EPE) & (EPE) & (EPE) & & (s) \\\hline\hline

  \multicolumn{6}{|c|}{Related works} \\\hline
  FlowNetS \cite{FlowNet}		& 2.71 & 4.50 &  5.45 & 38.5M	& 0.080	\\
  FlowNetC \cite{FlowNet}	 	& 2.19 & 4.31 &  5.87 & 39M	    & 0.150	\\
  SpyNet \cite{spynet}		& 2.63 & 4.12 &  5.57 & 1.2M	& 0.069	\\
  Zu et al. \cite{DenseF}	& 4.73 & -    & 10.07 & 2M		& 0.130	\\ 
  MSCSL	\cite{MSCSL}	    & 2.08 & 3.39 &  4.70 & -		& 0.060	\\\hline
  
  \multicolumn{6}{|c|}{Ours: base (stronger feature extractors)} \\\hline
  FlowNeXt32 & 2.58 & 3.82 & 4.38 & 33.8M & 0.066	\\
  FlowNeXt64 & 2.70 & 4.14 & 4.65 & 49.3M & 0.071	\\       %
  FlowNetRes & 2.26 & 3.56 & 4.06 & 52.2M & 0.055	\\
  FlowNetInc & 2.27 & 3.94 & 4.40 & 25.9M & 0.044	\\\hline %
  
  \multicolumn{6}{|c|}{Ours: with attention} \\\hline
  AttFlowNeXt32  & 1.94 & 5.08 & 6.91 & 26.8M & 0.060	\\
  AttFlowNeXt64  & 1.62 & 4.71 & 6.65 & 42.3M & 0.074	\\\hline
  
  \multicolumn{6}{|c|}{Ours: with midway} \\\hline
  MidFlowNeXt32 & 1.74 & 5.85  & 7.86 & 51.2M & 0.055	\\ %
  MidFlowNeXt64 & 1.48 & 5.30  & 7.45 & 66.8M & 0.076	\\\hline
    
  \multicolumn{6}{|c|}{Ours: with attention and midway} \\\hline
  AttMidFlowNeXt32  & 1.51 & 4.89 & 6.62 & 48.4M & 0.074	\\
  AttMidFlowNeXt64  & 1.60 & 5.25 & 6.74 & 64.0M & 0.092	\\\hline
  \end{tabular}
  \label{table:optical_flow_results}

\end{table}

\subsection{Optical flow estimation evaluation}
Table \ref{table:optical_flow_results} shows the optical flow estimation results. We evaluated our networks with the FlyingChairs dataset \cite{FlowNet}, and the MPI\_Sintel dataset \cite{sintel}.
FlyingChairs \cite{FlowNet} is a synthetic dataset designed specifically for training CNNs to estimate optical flow. It is created by applying affine transformations to real images and synthetically rendered chairs. The dataset contains 22,872 image pairs: 22,232 training and 640 test samples according to the standard evaluation split, with resolution 512$\times$384. MPI\_Sintel \cite{sintel} is a synthetic dataset derived from a short open source animated 3D movie. There are 1,628 frames, 1,064 for training and 564 for testing with a resolution of 1024$\times$436. The difference between them is that \textit{Clean pass} contains realistic illuminations and reflections, and \textit{Final pass} added additionally rendering effects like atmospheric effects and defocusing blurs. 

We measured the performance of our estimation networks via End Point Error (EPE). EPE is calculated using Euclidean distance between the ground truth optical flow and the estimated optical flow. For training, we used exclusively FlyingChairs \textit{train} set, since MPI\_Sintel is too small for large-scale deep learning training. Then, FlyingChairs is tested in its \textit{test} set, and both Sintel results are on their respective \textit{train} sets. 

\subsubsection{Discussion} \hfill 

\textbf{Stronger feature extractors generalize better}: Our base networks with stronger feature extractors perform similarly to FlowNetS and FlowNetC on our training dataset, FlyingChairs, but outperform both of these networks on both of the Sintel passes. This is because, while FlowNetS and FlowNetC achieved a good EPE on FlyingChairs, they do not generalize well to other datasets.

\textbf{Stronger feature extractors are more robust to noise}: Although some related works, e.g. MSCSL, achieved better \textit{Sintel clean} results, we observed that our base networks achieved better \textit{Sintel final} results. This proves that by adapting stronger feature extractors onto learning optical flow, the networks can understand motion better and ignore the additional atmospheric effects and motion blurs included in the \textit{final pass} better, thus obtaining closer results between the \textit{Sintel clean} and \textit{Sintel final} benchmarks. 

\textbf{Our method outperforms others in FlyingChairs}: By adding the attention mechanism, midway estimations and attention-midway combined, we significantly outperformed other methods on the training set, FlyingChairs, with an improvement of between 20\% to 45\%. However, we performed poorer on the Sintel dataset. We think it is because of the dataset difference between Sintel and FlyingChairs, also discussed in \cite{FlowNet}. Sintel has larger motions than FlyingChairs. When performing deconvolutions, the larger motions will more likely be upsampled less accurately, therefore resulting in an larger EPE. Hence, when we reached very accurate FlyingChairs estimations, which feature more smaller motions, we will suffer on Sintel. 
Nevertheless, our attention and midway networks, individually and combined, achieve the highest accuracy to date to our knowledge on the FlyingChairs dataset.

\textbf{Our method improves the speed-accuracy trade-off}: Adding an attention module improves runtime (faster execution) and reduces the model size (less memory consumed). Then, the midway module reduces the runtime but increases the model size. The best speed-accuracy trade-off is achieved on the FlyingChairs dataset by combining attention and midway since, although the runtime increases by 10\%, the accuracy increases by 45\%.

\subsection{Gesture recognition evaluation}

Table \ref{table:gesture_results} shows the gesture recognition accuracy when using our optical flow estimation methods. For that, we used our MIBURI dataset and the Isolated Gesture Recognition Challenge (IGR) dataset. In MIBURI, our training split contains the first three gesture repetitions of each user for training, and the remaining for testing. The IGR dataset (in ICPR 2016) was derived from one of the popular gesture recognition datasets, the ChaLearn Gesture Dataset 2011 \cite{CHALEARN}. It includes in total 47933 RGB-D gesture videos, where each RGB-D video represents an isolated gesture (we only used the RGB channels). The dataset is split in 35878 videos for training, 5784 videos for validation and 6271 videos for testing. The resolution is 240$\times$320. There are 249 gestures labels performed by 21 different individuals. The gestures are very diverse (e.g., music notes, diving signals, referee signals), which makes it more challenging than MIBURI.

We chose the optical flow branch of the Temporal Segment Network (TSN) \cite{TSN} as our gesture recognition method, since its performance is state-of-the-art and the speed-accuracy trade-off is well balanced. We compared it to three baselines: First, improved dense trajectories (iDT) \cite{idt} with Fisher vector (FV) encoding \cite{fv} and support vector machines (SVM) classifier, which combined were the state of the art before deep learning \cite{idt}. Second, TSN (default) uses traditional (i.e., non-deep learning-based) optical flow estimation \cite{TVL1}. Third, TSN/FlowNetS is TSN's optical flow branch replaced with optical flow estimation from FlowNetS \cite{FlowNet}.

\noindent
\begin{table}[t]
	\setlength{\tabcolsep}{4pt}
	\centering\fontsize{8}{10}\selectfont
	\bigskip\caption{Gesture recognition accuracy with our proposed optical flow. The runtime reported is on MIBURI, and includes gesture recognition optical flow estimation.}
	\label{table:results_TSN}
	\begin{tabular}{|c|c|c|c|c|}
		\hline
		\multirow{2}{*}{Model} & MIBURI & ChaLearn IGR & Recognition \\
		& accuracy (\%) & accuracy (\%) & runtime (s)          \\\hline\hline
		
		\multicolumn{4}{|c|}{Baseline}            \\ \hline
		iDT+FV+SVM        & 98.3 & - & 27.1         \\
		TSN (default)     & 96.9 & 94.55 & 11.1         \\
		TSN/FlowNetS 	  & 73.9 & 84.98 & 1.7          \\\hline

		\multicolumn{4}{|c|}{Ours: base (stronger feature extractors)}      \\\hline
		TSN/FlowNeXt32 & 92.1 & 76.83 & 4.7      \\
		TSN/FlowNeXt64 & 93.8 & 81.79 & 7.9      \\
		\hline
		
		\multicolumn{4}{|c|}{Ours: with attention} \\\hline
		TSN/AttFlowNeXt32 & 96.9 & 84.31 & 4.1    \\
		TSN/AttFlowNeXt64 & 96.9 & 87.51 & 6.4    \\
		\hline
		
		\multicolumn{4}{|c|}{Ours: with midway} \\\hline
		TSN/MidFlowNeXt32 & 96.9 & 85.94 & 6.0 \\ %
		TSN/MidFlowNeXt64 & 96.9 & 85.02 & 8.5 \\
		\hline

		\multicolumn{4}{|c|}{Ours: with attention and midway} \\\hline
		TSN/AttMidFlowNeXt32 & 96.9 & 86.00 & 7.2 \\
		TSN/AttMidFlowNeXt64 & 98.4 & 85.43 & 9.6 \\
		\hline

	\end{tabular}
	\label{table:gesture_results}
\end{table}

\subsubsection{Discussion} \hfill

\textbf{Attention improves gesture recognition}: Adding attention to our deep learning-based networks translated to a better gesture recognition performance in both datasets. This effect can be observed for TSN/FlowNeXt\{32/64\} vs TSN/AttFlowNeXt\{32/64\} and TSN/MidFlowNeXt\{32/64\} vs TSN/AttMidFlowNext\{32/64\}. As we hypothesized, since attention provides a more refined human silhouette, human motion is better represented, and thus, the performance of optical flow-based gesture recognition methods is improved.

\textbf{Better EPE does not necessarily imply better gesture recognition}: In our experiments with MIBURI, improvement in optical flow and gesture recognition is correlated. For ChaLearn IGR on the other hand, in spite of being outperformed in optical flow estimation by our networks, TSN/FlowNetS achieves better recognition accuracy than our base.
The reason for this mismatch is that, whereas only correctly-recognized motions account for gesture recognition, EPE metric also takes into account correctly-recognized background pixels.
The way of obtaining a good score in both metrics is by providing an accurate representation of the human silhouette, like in our attention mechanism.

\textbf{Our method improves the speed-accuracy trade-off}: Compared to traditional optical flow estimation methods, deep learning-based methods allow for a faster gesture recognition, which has a lot of impact in HRI. With our MIBURI dataset, comparing TSN/FlowNeXt\{32/64\} vs TSN/AttFlowNeXt\{32/64\}, adding attention improved recognition accuracy by 7\% to 10\% while runtime decreases, achieving the best speed-accuracy trade-off. While other configurations achieve higher accuracies, their recognition runtime makes them prohibitive for HRI.

\section{Conclusion}
In this paper, we presented a novel optical flow estimation method and evaluated it in a gesture recognition pipeline for human-robot interaction.
We presented four ways of improving optical flow estimation: stronger feature extractors, attention mechanism, midway mechanism, and a combination of the latter. Due to the lack of gesture datasets for commanding house service robots, we generated a gesture dataset called MIBURI. We proved that our method allows for a better representation of motion, improving the optical flow estimation of existing deep learning-based gesture recognition work. This improvements also lead to more accurate gesture recognition, and achieve a better speed-accuracy trade-off. In the future, we plan to extend this work by integrating it with a robot, and comparing the efficiency of gesture-based human-robot interaction against current voice-controlled approaches.

\section*{ACKNOWLEDGMENT}
This work was partially supported by JST CREST Grant Number JPMJCR1403, Japan.
The authors have participated in the HSR developer community\footnote{https://newsroom.toyota.co.jp/jp/detail/8709536} and made use of HSR hardware and software platforms.

\bibliographystyle{IEEEtran}
\bibliography{Bibliography}

\end{document}